# Void-Defect Induced Magnetism and Structure Change of Carbon Material-II: Graphene Molecules

Norio Ota[1], Aigen Li[2], Laszlo Nemes[3] and Masaaki Otsuka[4]

[1]Graduate school of Pure and Applied Sciences, University of Tsukuba, *1-1-1 Tennodai, Tsukuba-City Ibaraki, 305-8571, Japan*
[2]Department of Physics and Astronomy, University of Missouri, *Columbia, MO 65211, USA,*
[3]Research Center for Natural Sciences, Ötvös Lóránd Research Network, *Budapest 1519, Hungary*
[4]Okayama Observatory, Kyoto University, *Asakuchi Okayama, 719-0232, Japan,*

Void-defect is a possible origin of ferromagnetic feature on pure carbon materials. In our previous paper, void-defect on graphene-nanoribbon show highly polarized spin configuration. In this paper, we studied cases for graphene molecules by quantum theory, by astronomical observation and by laboratory experiment. Model molecules for the density functional theory are graphene molecules of $C_{23}$ and $C_{53}$ induced by a void-defect. They have carbon pentagon ring within a hexagon network. Single void has three radical carbons, holding six spins. Those spins make several spin-states, which affects to molecular structure and molecular vibration, finally to infrared spectrum. The stable spin state was triplet, not singlet. This suggests magnetic pure carbon molecule. It was a surprise that those molecules show close infrared spectrum with astronomically observed one, especially observed on carbon rich planetary nebulae. We could assign major band at 18.9 micrometer, and sub-bands at 6.6, 7.0, 7.6, 8.1, 8.5, 9.0 and 17.4 micrometer. Also, calculated spectrum roughly coincides with that of laboratory experiment by the laser-induced carbon plasma, which is an analogy of cosmic carbon creation in interstellar space.
[To be published on Journal of the Magnetics Society of Japan (2021), e-mail to Norio Ota: n-otajitaku@nifty.com ]

## 1. Introduction

Graphene and graphite like carbon materials are candidates for showing ferromagnetic like hysteresis[1)-6)]. There are many capable explanations based on impurities[7)], edge irregularities[8)-10)] or defects[11)-16)]. The density functional theory (DFT) shows good coincidence with experiments[17)-19)] and revealed reasonable consistency with several fundamental theories[20)-21)]. Despite such many efforts, origin of magnetic ordering could not be thoroughly understood. Our previous study on graphene nano-ribbon (GNR) revealed that void-defect brings unusual highly polarized spin configuration and structure change, which will be opened in this journal under the same title as number -I. In this paper, we like to apply our study to graphene molecules, smaller than GNR. Unfortunately, on laboratory experiment, small pure carbon molecule does not show any magnetic feature. The reason may be molecule-to-molecule interaction at high density $(10^{10} \sim 10^{23}$ molecules/cm$^3)$ conditions on earth. Radical carbon's magnetic feature in small molecules may be cancelled out by complex interaction between molecules. Now, we should look at astronomical carbon dust floating in interstellar and circumstellar space. Molecules are kept under peculiar condition of ultra-low density (1~100 molecules/cm$^3$), almost no interaction each other. Molecular structure affects molecular vibration and infrared spectrum. In this study, calculated infrared spectrum will be compared with the astronomically observed one, also with the laboratory experiment of the laser induced carbon plasma.

Fullerene $C_{60}$ was discovered by Kroto et al.[22)] (the 1996 Nobel Prize) in the sooty residues of vaporized carbon. They already suggested that "fullerene may be widely distributed in the universe". The presence of $C_{60}$ in astrophysical environments was revealed by the detection of a set of emission bands at 7.0, 8.45, 17.3 and 18.9 μm[23)-27)]. Typical astronomical objects are the Galactic planetary nebula (PNe) Tc1[23)] and the Small Magellanic Cloud Lin49[28)]. Observed spectra were compared with experiment and theory[23)29)-31)]. However, there remain undetermined observed bands, not be explained by $C_{60}$.

It is well known that graphene is a raw material for synthesizing fullerene[32)33)]. By observation of Lin49, Otsuka et al.[28)] suggested the presence of small graphene. Graphene was first experimentally synthesized by Geim and Novoselov[34)] (the 2010 Nobel Prize). The possible presence of graphene in space was reported by Garcia-Hernandez et al.[35)-37)]. These infrared features appear to be coincident with planar $C_{24}$ having seven carbon hexagon rings. However, full observed bands still cannot be explained by $C_{24}$ or hexagon network molecules[38)]. Some hints come from carbon SP3 defect among SP2 network caused by single void-defect in carbon hexagon network by Ota[39)].

Also, Galue & Leines[40] predicted the physical model supposing pi-electron irregularity. In this paper, we apply one assumption of single void-defect on graphene molecule. At the first part, DFT calculation will be applied to model molecules. Void position discriminates detailed molecule species. Spin dependent calculation will be done to obtain detailed molecular structure and molecular vibrational infrared spectrum. At the second part, we will compare such calculated spectra with astronomically observed one. Finally, we like to compare with laboratory experiment of laser induced carbon plasma spectrum[41)42)], which will be an analogy of cosmic carbon creation in space.

## 2. Model Molecules and Calculation Method

Model molecules are illustrated in Fig. 1 starting from $C_{24}$ having seven carbon hexagon rings[35]. To find size dependence, we add larger one of $C_{54}$. In this paper, we apply one assumption of single void-defect on such initial molecules. In laboratory experiment, high speed particle can create such void. Similarly, in astronomical space, the cosmic ray as like proton may attack graphene[35-37]. As illustrated on top left of Fig. 1, high speed particle attacks graphene and kick out one carbon atom. DFT calculation shows molecular configuration of void-defect induced graphene molecules as like $C_{23}$, and $C_{53}$, which have one pentagon ring among hexagon networks. Molecular structure depends on the void position as marked by a, b, c, d, e, f, in left of Fig. 1. Void induced molecule species is named by suffixing a, b, and so on, such as ($C_{23}$-a) and ($C_{23}$-b). Molecular energy of ($C_{23}$-a) is 1.04 eV higher than that of ($C_{23}$-b). In this study, we supposed isolate molecule, which means no molecule-to-molecule interaction, and no energy competition between different void induced species. While we need energy competition between spin states for every individual species. It should be noted that one void has 3 radical carbons and holds 6 spins. Such spins bring spin-multiplicity, which affect to molecular structure and molecular vibration, finally to infrared spectrum.

In calculation, we used DFT[43) 44)] with the unrestricted B3LYP functional[45]. We utilized the Gaussian09 software package[46] employing an atomic orbital 6-31G basis set[47]. Unrestricted DFT calculation was done to have the spin dependent atomic structure. The required convergence of the root-mean-square density matrix was $10^{-8}$. Based on such optimized molecular configuration, fundamental vibrational modes were calculated, such as C-C stretching modes, C-C bending modes and so on, using the Gaussian09 software package. This calculation also gives harmonic vibrational frequency and intensity in infrared region. The standard scaling is applied to the frequencies by employing a scale factor of 0.975 for pure carbon system taken from the laboratory experimental value of 0.965 based on coronene molecule of $C_{24}H_{12}$[39]. Correction due to anharmonicity was not applied to avoid uncertain fitting parameters. To each spectral line, we assigned a Gaussian profile with a full width at half maximum (FWHM) of 4cm$^{-1}$.

## 3. Spin State Analysis and Structure Change

We tried the multiple spin-state analysis. Example is shown in Fig. 2 for spin state of $S_z$=2/2 of ($C_{23}$-a) and ($C_{23}$-b). In this study, we dealt total molecular spin **S** (vector). Molecule is rotatable material, easily follows to the external magnetic field of z-direction. Projected component to z-direction is maximum value of $S_z$, which is good quantum parameter. In molecular magnetism, $S_z$=2/2 is named as triplet spin-state. Initial void-defect holds 3 radical carbons and allows 6 spins as illustrated in left. Six spins make capable spin-states of $S_z$=0/2, 2/2, 4/2 and 6/2. Among them, calculated energy of $S_z$=4/2 and 6/2 resulted unstable high energy. Here, we like to compare cases of $S_z$=0/2 and 2/2. One radical carbon holds two spins, which are forced to be parallel up-up spins (by red arrows in Fig. 2) or down-down spins (blue) for avoiding large coulomb repulsion due to Hund's rule[58]. Spin alignment model is illustrated on middle. Three couples of spin-pair will be partially cancelled and remain one pair to be $S_z$=2/2. Such speculation is confirmed by DFT calculation. Spin cloud is mapped on right at a cutting surface of spin density at 10e/nm$^3$. We can see up-spin major spin cloud coincident with above simple speculation.

As shown in panel of ($C_{23}$-a) in Fig. 1, molecular structure of $S_z$=0/2 shows bending top carbon (see a blue circle of side view), whereas $S_z$=2/2 shows flat one as illustrated on right column. It should be noted that stable spin state is $S_z$=2/2, not $S_z$=0/2. Molecular energy of $S_z$=2/2 was 0.49 eV lower (stable) than that of $S_z$=0/2. Cause of bending carbon of $S_z$=0/2 comes from partial inclusion of SP3 component among SP2 network[39)40)]. Such inclusion increases total molecular energy. Similar result was confirmed again for a case of ($C_{23}$-b).

Other molecules' structure and energy are listed in Fig. 1. Energy change from the spin state $S_z$=0/2 to the $S_z$=2/2 was noted by ΔE. Stable spin-state is framed by bold green. In most species, $S_z$=2/2 is stable (lower energy) than $S_z$=0/2. Exception is only ($C_{53}$-a) to show stable state of $S_z$=0/2. This comes from complete structure change, that is, void at a-site induces SP3-bond among SP2-networks as marked by a red circle in a column of ($C_{53}$-a, $S_z$=0/2).

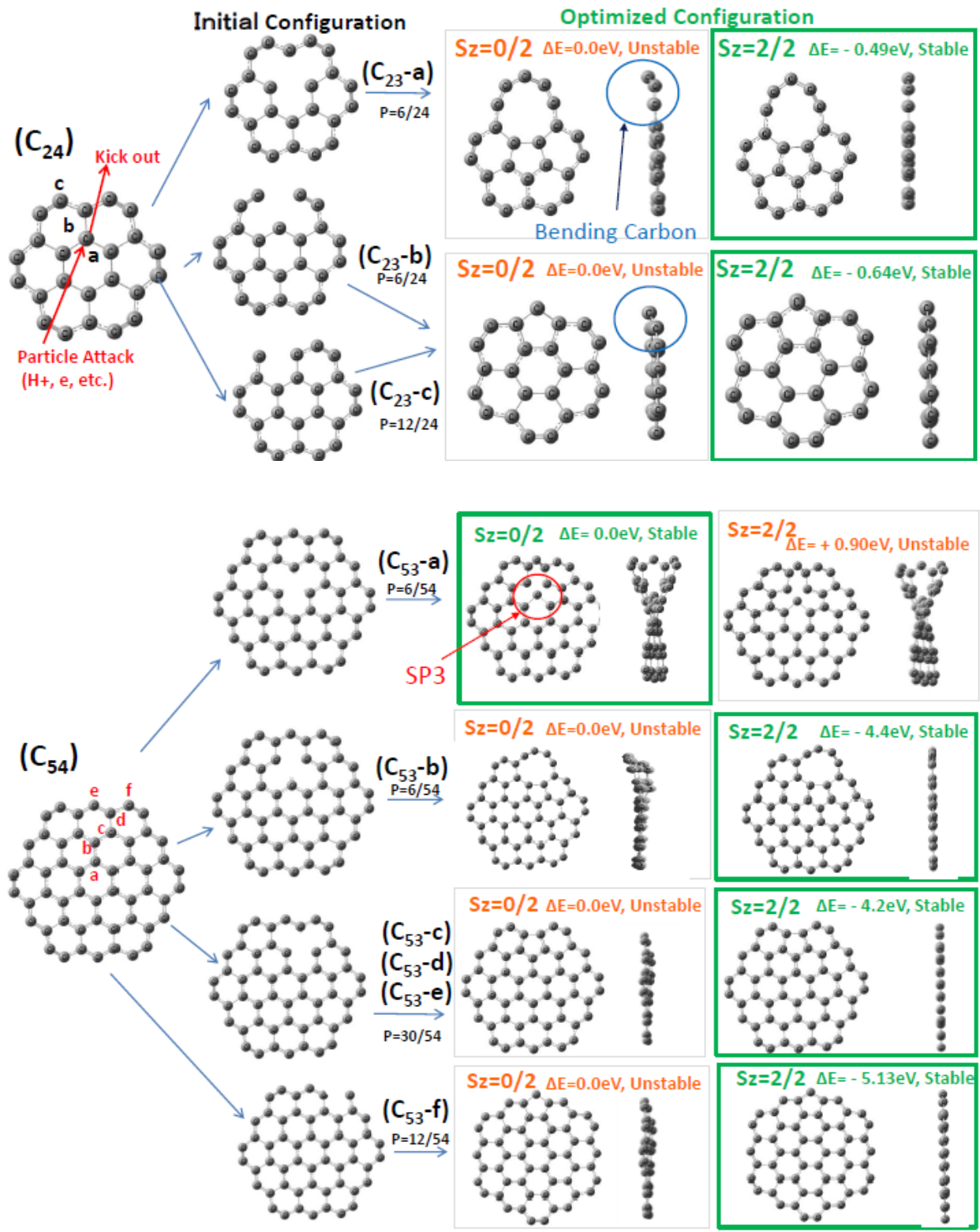

**Fig. 1** Model graphene molecules. Void-defect positions are classified by suffix a, b, and so on. Calculated molecular structures are illustrated for both spin-states of $S_z=0/2$ and $S_z=2/2$. Total energy difference ΔE was noted in every column. Stable spin-state was framed by bold green.

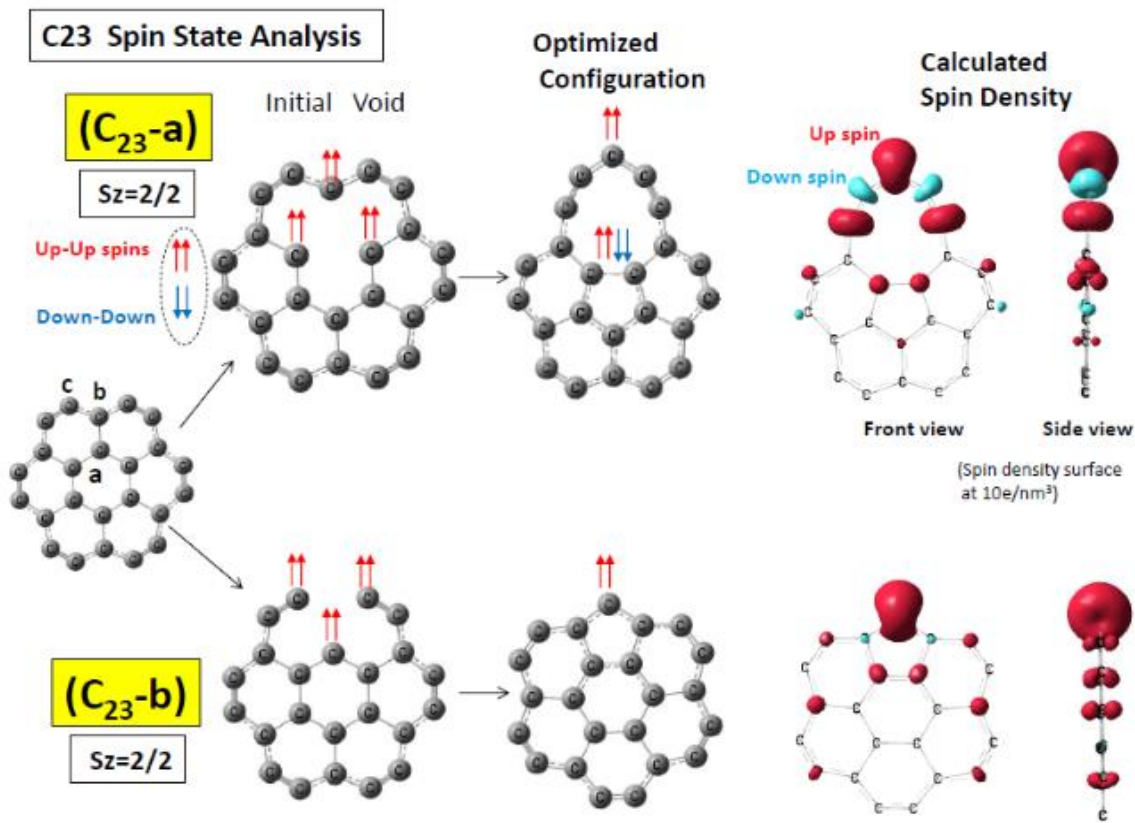

**Fig. 2** Spin alignment for ($C_{23}$-a) and ($C_{23}$-b) with spin state of $S_z$=2/2. Initial void holds 3 radical carbons and allow 6 spins. By re-combination of carbon atoms, there arises partial cancelling of (up-up) spins by (down-down) one. DFT calculated spin density of $S_z$=2/2 is illustrated on right.

### 4. Calculated Infrared Spectrum

Calculated infrared spectra are shown in Fig. 3. Left columns are cases for spin state $S_z$=0/2, while right one $S_z$=2/2. Spectrum of stable spin state was framed by bold green. Astronomically observed major emission line of 18.9 μm was marked by a green dotted line. In case of ($C_{23}$-a), we can see one carbon pentagon ring and an unusual 9 membered ring. Calculated infrared spectrum for $S_z$=2/2 show 18.8 μm peak close to observed band of 18.9μm. The species of ($C_{23}$-b, $S_z$=2/2) [same structure for ($C_{23}$-c)] have one pentagon ring and show calculated twin major bands at 18.9 and 19.1 μm.

In case of ($C_{53}$-a), stable spin state was $S_z$=0/2, of which spectrum show peaks at 21.7 μm and 19.5 μm to be far from observed one. The spectrum of ($C_{53}$-b, $S_z$=2/2) show a peak at 19.0 μm close to observed one. Also, ($C_{53}$-c, $S_z$=2/2) [same for ($C_{53}$-d) and ($C_{53}$-e)] demonstrates major band at 18.9 μm, just the same with observed one. In case of ($C_{53}$-f, $S_z$=2/2), we can see twin bands at 18.4 and 18.9 μm. Thus, we could obtain several species suitable for assigning the astronomical observation.

### 5. Fundamental Mode Analysis

Fundamental vibrational mode of ($C_{23}$-b, $S_z$=2/2) was analyzed and summarized in Table 1. There are 63 modes for 23 carbon atoms. Energy diagram is illustrated on top right of Table 1. Zero-point vibrational energy is 27634 cm$^{-1}$ (=3.42 eV). The lowest vibrational energy of Mode-1 is 94.9 cm$^{-1}$ (=108.06 μm, 0.012 eV). The highest Mode-63 is 2067 cm$^{-1}$ (=4.96 μm, 0.26 eV). Vibrational behaviors are noted on right column of the table. They are all carbon to carbon (C-C) in-plane stretching modes parallel to molecular plane. Combination of vibrating bonds is different for each mode. Bond is named by a, a', b, b' etc. illustrated on a molecule structure. For example, there are two calculated modes corresponding to observed 18.9 μm band. One is Mode-24 showing 18.8 μm in-plane C-C stretching at bonds of c, f, c', f', i, l, i' and l'. Another is Mode-23 showing 19.1 μm stretching at c, f, c', f', o, p, and o'. Other major modes will be discussed in next section by comparing astronomically observed bands.

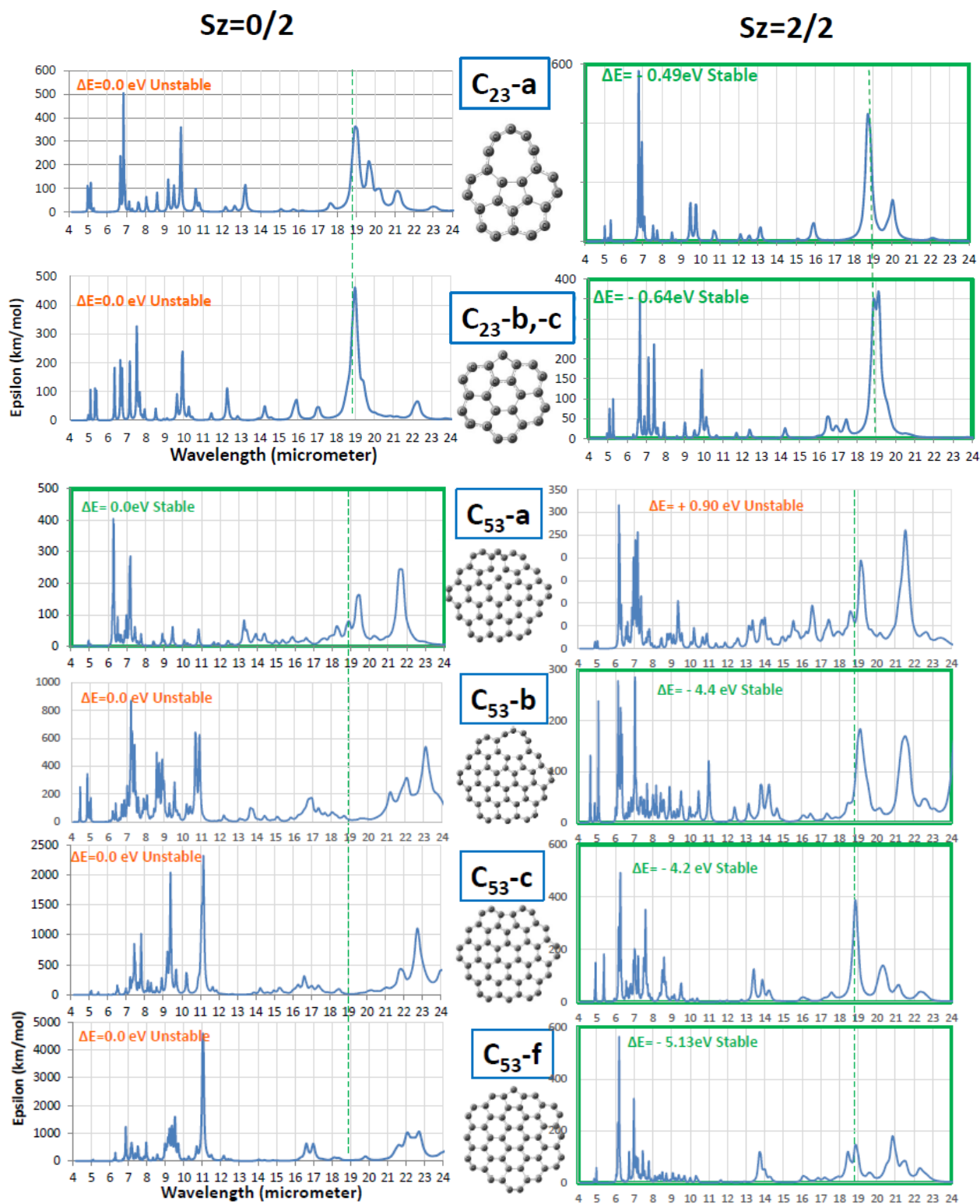

Fig. 3 Calculated infrared spectrum. Left figures are spectra for the spin state $S_z=0/2$, while right one for $S_z=2/2$. Stable spin state is framed by bold green.

**Table 1** Fundamental mode analysis of ($C_{23}$, $S_z=2/2$).

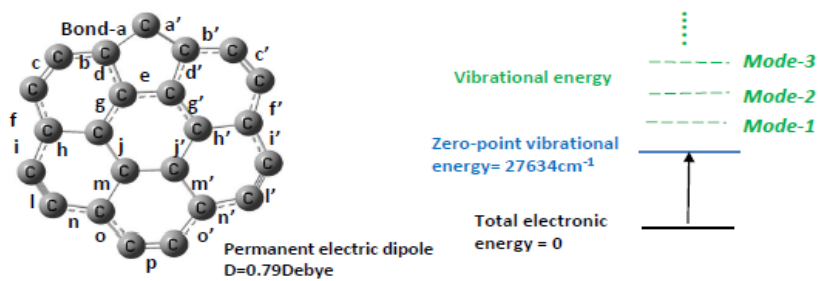

| Astronomically Observed Band (µm) | Experimental Band (laser induced carbon plasma) (µm) | Fundamental Mode of (C23-b, Sz=2/2) , DFT calculation ||||
|---|---|---|---|---|---|
| | | Mode number | Wavelength (µm) | Intensity (km/mol) | Vibrational Behavior, Carbon to carbon in-plane stretching, bond positions |
| 18.9µm | | Mode-23 <br> -24 | 19.1µm <br> 18.8 | 92.6 <br> 78.8 | c, f, c', f', i, l, i', l' <br> c, f, c', f', o, p, a o' |
| 17.4 | | -25 | 17.4 | 12.4 | c, f, c', f' |
| 16.7 | | -27 | 16.5 | 17.7 | b, b', c, c' |
| 10.0 | 10.2 | -42 | 9.9 | 52.6 | b, b', c, c' |
| 7.5 | 7.4 | -52 | 7.4 | 72.7 | a, a, a', b, b', g, g', j, j' |
| 7.0 | | -53 | 7.1 | 59.0 | a, a', b, b', d, d', g, g', j, j', o, o' |
| 6.6 | 6.7 | -55 <br> -56 | 6.7 <br> 6.6 | 94.0 <br> 40.5 | a, a', b, b', f, f', i, i' <br> n, n', o, o' |

## 6. Comparison with Astronomical Observation

### 6.1 Single molecule spectrum

On top of Fig. 4, astronomically observed infrared spectra are illustrated for carbon rich planetary nebulae of Lin49 (dark blue curve) and Tc1 (red one). In both cases, we can see major band at 18.9 µm, also a side band at 17.4 µm, and shorter wavelength bands at 6.6, 7.0, 7.5, 8.1 and 8.5 µm. Sharp atomic emission lines are marked by arrows of [NeII] and [SIII] for scaling observed wavelength. It should be noted that the observed spectra are seen in emission. A nearby star as like a central star of Tc1 nebula may illuminate the molecules and excites them to give rise infrared emission. Detailed discussion was done by Li and Drain[48),49)]. We regard here that DFT calculated absorbed spectrum is mirror image of emission one in case of sufficient large photon energy excitation due to the theory of Einstein's emission coefficient[48),49)].

It was amazing to see good coincidence with observed spectra and calculated one of ($C_{23}$-b, $S_z=2/2$). Detailed comparison is summarized in Table 1. Observed major band at 18.9 µm could be identified by two modes of Mode-23 (19.1 µm) and Mode-24 (18.8 µm). Also, observed 6.6 µm band was reproduced well by calculated two modes of Mode-55 (6.7 µm) and Mode-56 (6.6 µm). Again, observed bands at 7.0, 7.5, 10.0, 16.7, and 17.4 µm could be reproduced well respectively by Mode-53 (for 7.1 µm), Mode-52 (7.4 µm), Mode-42 (9.9 µm), Mode-27 (16.5 µm), and Mode-25 (17.4 µm).

It was amazing again that large molecule cases as like ($C_{53}$-c, $S_z=2/2$) show good coincidence with observation as compared on a bottom panel of Fig. 4. It was suggested that there may be little size dependence. Void-defect induced graphene molecules generally contribute on cosmic spectra.

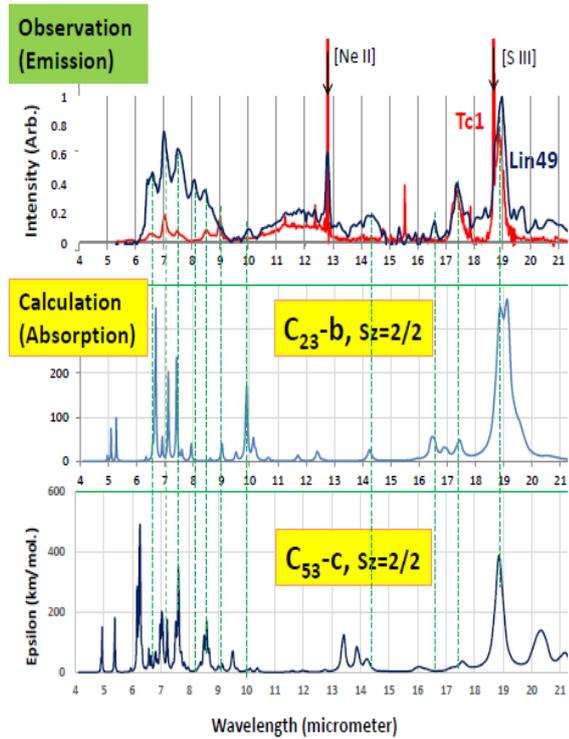

**Fig. 4** Astronomically observed emission spectrum for planetary nebula Tc1 (on top, by red) and Lin49 (black) compared with calculated spectrum of ($C_{23}$-b, $S_z$=2/2), and ($C_{53}$-c, $S_z$=2/2).

### 6.2 Weighting sum spectrum

We like to know total contribution of all model molecules, not depending on peculiar molecular species. Favorable way is the weighting sum method. In case of $C_{23}$-family, void-a species has 6 identical positions among 24 carbon sites of $C_{24}$ mother molecule. Weighting sum coefficient-p should be p=6/24. Also, coefficient of void-b species is p'=6/24 and void-c p"=12/24. Weighting sum of epsilon $\varepsilon$($C_{23}$-family) will be obtained by the following equation (1).

$\varepsilon$($C_{23}$-family)
 = p·$\varepsilon$($C_{23}$-a, $S_z$=2/2)+
   p'·$\varepsilon$($C_{23}$-b, $S_z$=2/2)+
   p"·$\varepsilon$($C_{23}$-c, $S_z$=2/2))
 = (6/24)·$\varepsilon$($C_{23}$-a, $S_z$=2/2)+
   (6/24)·$\varepsilon$($C_{23}$-b, $S_z$=2/2)+
   (12/24)·$\varepsilon$($C_{23}$-c, $S_z$=2/2))-------(1)

Weighting sum spectrum of $C_{23}$-family was illustrated on middle of Fig. 5. It was amazing that we can see good coincidence between observation and calculation. In case of larger molecule of $C_{53}$, weighting sum of $C_{53}$-family could well reproduce observed one as shown on bottom of Fig. 5, where void creation parameter p was noted in Fig. 1.

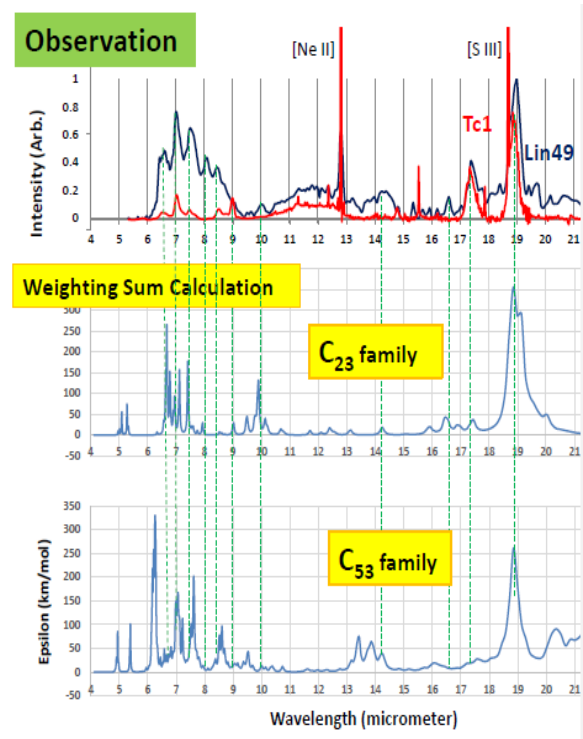

**Fig. 5** Weighting sum spectra for $C_{23}$-family and $C_{53}$-family. Void creation capability parameter-p of each molecule is considered as weighting sum coefficient as noted in Fig. 1.

### 7. Comparison with Laboratory Experiment

Nemes et al.[41)42)] did the laser induced carbon plasma experiment as an analogy of carbon dust creation in interstellar space. Bulk graphite was heated and excited by Nd:YAG laser with wavelength of 1.064 μm (1.16 eV) in atmospheric pressure Argon gas. Evaporated carbon molecules emit infrared light. The emission was recorded by an HgCdTe detector.

Several theories dealt with the carbon dust creatin in space[50)-57)]. Calculated average size of carbon dust is about 1nm, which is similar size with $C_{23}$ and $C_{53}$ in this study. Temperature of star explosion gas will be 2000 K after 300 days of the explosion, and finally cooled as dust cloud[57)]. In case of laser induced plasma experiment, excited carbon temperature will be 4500 to 7000 K, and finally cooled to room temperature[42)].

Experimental spectrum is shown in panel (A) of Fig. 6. We compared with weighting sum spectrum of $C_{23}$-family and $C_{53}$-family in panel (B). The experimental 7.4 μm peak could be reproduced by calculated 7.5 μm band of $C_{23}$-family. We can see plateau from 6 to 7 μm, which may correspond to calculated bands at 6.7 and 7.2 μm. At a range of 5 μm, we can suggest some contribution by both $C_{23}$-family and $C_{53}$-family. Also, at a range of 10μm, $C_{23}$-family may contribute. It should be noted that void-defect induced graphene could roughly explain laboratory experiment.

Comparison to astronomical observation was shown in panel (C). Astronomically observed bands are featured by black dotted lines. We can find again good reproducibility by calculation.

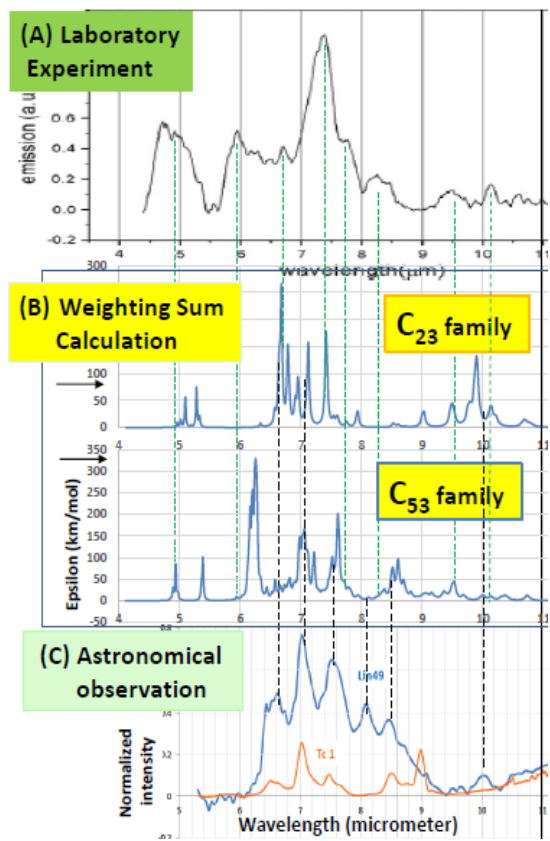

**Fig. 6** (A) Infrared spectrum of the laser induced carbon plasma, (B) calculation for $C_{23}$-family and $C_{53}$-family, (C) astronomically observed spectra.

## 8. Conclusion

Void-defect is a possible origin of ferromagnetic like feature of graphene and graphite. In this paper, graphene molecules were studied.
(1) Starting model molecules were graphene molecules of $C_{24}$ and $C_{54}$ for the density functional theory analysis. Single void-defect creates $C_{23}$, and $C_{53}$ having carbon pentagon ring among hexagon networks.
(2) Single void holds six spins to bring spin multiplicity. DFT calculation show the most stable spin-state to be $S_z=2/2$, not $S_z=0/2$.
(3) We compared calculated infrared spectrum with astronomically observed one. Interstellar carbon is expected to avoid complex molecule-to-molecule interaction. The triplet spin state of $C_{23}$ and $C_{53}$ could explain astronomically observed spectra of Tc1 and Lin49 nebulae for a major band at 18.9μm and sub-bands from 6 to 18μm.
(4) By a weighting sum method applied to different void position species, again we could reproduce well observed infrared spectra.
(5) The laboratory experiment on laser induced carbon plasma was tried as an analogy for carbon dust creation in space. Experimental infrared emission spectra could be roughly explained by the weighting sum spectra of $C_{23}$ and $C_{53}$.

## Acknowledgement

Aigen Li is supported in part by NSF AST-1311804 and NASA NNX14AF68G.